\documentclass[pra]{revtex4}
\usepackage{amsmath}
\usepackage{graphicx}
\begin{document}
\title{Intermittency and Slip-Size Distribution in the Block-Spring Model of Earthquakes}
\author{Hidetsugu Sakaguchi and Shingo Morita}
\affiliation{Department of Applied Science for Electronics and Materials,
Interdisciplinary Graduate School of Engineering Sciences, Kyushu
University, Kasuga, Fukuoka 816-8580, Japan}
\begin{abstract}
The Carlson-Langer model is a deterministic model of earthquakes. There were many investigations of this model, but its complicated spatio-temporal dynamics is not yet completely understood. We again study the model equation numerically, and obtain two new results. Firstly, we try to understand the complicated dynamics from the viewpoint of spatio-temporal intermittency. We study the intermittent dynamics of a two-block system  using a one-dimensional map, which is useful for understanding an intermittent spatio-temporal chaos in larger systems. 
Large slips occur intermittently after repeated small slips when the parameter of the velocity weakening friction is large.  Small slips reduce the spatial irregularity of block displacement, and large slips occur on a relatively flat configuration of blocks. Secondly, we study the effect of spatial heterogeneity such as asperity using percolation clusters. We show the numerical results of the slip-size distribution in the block-spring model on percolation clusters, and find that the $b$ parameter for the magnitude distribution is close to 2/3 when the velocity weakening parameter is intermediate. 
\end{abstract}
\maketitle
\section{Introduction}
Earthquakes are research subjects in seismology and geophysics; however, their statistical properties and complex dynamics have also been studied in statistical physics and nonlinear physics. Earthquakes exhibit various types of power laws, the most well-known of which is the Gutenberg-Richter law, which states  that  the energy released during earthquakes obeys a power law.~\cite{rf:1,rf:2} 
 
Burridge and Knopoff proposed a block-spring model of earthquakes.~\cite{rf:3}
However, they could not reproduce the power law in their numerical simulation.
Later, Carlson et al. used a type of velocity-weakening friction and observed a power law in the range of small magnitude in a homogeneous block-spring model.~\cite{rf:4,rf:5} Since then, many authors have intensively studied the Carlson-Langer model and its modification. Cao and Aki studied a block-spring model using a state-dependent friction law.~\cite{rf:6} Sakaguchi found  phase transition in a block-spring model pulled at one end, using a different type of state-dependent friction.~\cite{rf:7} Carlson studied a two-dimensional version of the Carlson-Langer model.~\cite{rf:8}
de Sousa Vieira et al. pointed out the transition of the magnitude distribution by changing the parameter of the velocity-weakening friction.~\cite{rf:9} 
Mori and Kawamura studied a similar transition of the magnitude distribution in two-dimensional block-spring models.~\cite{rf:10} Moreover, Nakanishi proposed a simpler cellular automaton model.~\cite{rf:11}

The power law behavior found in the Carlson-Langer model might be related to the notion of the self-organized critical phenomenon. Bak and Tang and Olami et al. proposed cellular automaton models related to earthquakes, which exhibit the self-organized criticality.~\cite{rf:12,rf:13}
In these models, the crust on which earthquakes occur is considered to be in a self-organized critical state, and the power laws are considered to originate from the critical phenomena.
On the other hand, there are several models that show that the power laws originate from the fractal geometry of faults or asperities. Earthquakes occur on active faults. Fractal geometry was studied for actual faults such as the San Andreas Faults.~\cite{rf:14} The position where two plates are firmly stuck is called an ``asperity''.  There are various sizes of asperities, and the slip at a large asperity is considered to generate a large earthquake.    
The size distribution of earthquakes is explained using such fractal geometry by several authors.~\cite{rf:15} Ohtsuka proposed a ``Go-Game''model of earthquakes as a fractal model, which is equivalent to a percolation model.~\cite{rf:16,rf:17}  There is also an intermediate model such as a cellular automaton model with fractal heterogeneity.~\cite{rf:18}  
These two types of model are further related to the controversy of whether or not earthquakes can be predicted.~\cite{rf:19, rf:20}  

Although many authors have studied the Carlson-Langer model, its complex dynamics is not yet completely understood. In this paper, we study the model again and obtain two new results. Firstly, we try to understand the complicated dynamics from the viewpoint of spatio-temporal intermittency. 
 We first study the intermittent dynamics of a two-coupled block-spring system using a one-dimensional map. 
There have been several investigations of one-block and two block systems.~\cite{rf:21,rf:22,rf:23} 
Our system is almost the same as such systems, but here we pay attention to the intermittent dynamics of a large slip following many small slips, and elucidate the mechanism of the intermittency. Then we consider spatio-temporal chaos in one-dimensional systems. The spatial irregularity is reduced during the regime of small slips and then a large slip occurs on a relatively flat configuration of blocks. On the other hand, the spatial irregularity is amplified during large slips, and small slips occur on the rough configuration of blocks after a large slip. A related argument was given in the original paper.~\cite{rf:4} However, here we will clarify it by using the result of the two-block system and by measuring the irregularity in one-dimensional systems.  The intermittent dynamics is discussed in Sects.~2-4.
  Secondly, we study the block-spring system on percolation clusters to incorporate the spatial inhomogeneity such as the asperity, and show that the exponent of the power law is about $2/3$, which is close to the exponent in the original Gutenberg-Richter law~\cite{rf:2}. The ``Go-Game'' model by Ohtsuka is a static model, and the numerical investigation of the block-spring models on percolation clusters has not yet been reported. The power law of the slip-size distribution is discussed in Sects.~5 and 6.

\section{Carlson-Langer Model and Stick-Slip Dynamics in One-Block Systems} 
The Carlson-Langer model is a block-spring model. 
In a one-dimensional model, $N$ blocks of mass $m=1$ are coupled with nearest-neighbor blocks with a spring of spring constant $k$. Each block is further coupled to a rigid pulling plate moving with a constant velocity $v_0$.   
The one-dimensional Carlson-Langer model is expressed as 
\begin{equation}
\frac{d^2x_i}{dt^2}=k(x_{i+1}-2x_i+x_{i-1})+(v_0t-x_i)-\phi(v_i)
\end{equation}
where $v_i$ denotes the velocity of the block $dx_i/dt$,  and $\phi(v_i)$ represents the kinetic friction. The maximum static friction is set to 1. In the Carlson-Langer model, the kinetic friction $\phi(v)$ is assumed to be 
\begin{equation}
\phi(v)=\frac{1-\sigma}{1+2\alpha v/(1-\sigma)}
\end{equation}
for $v>0$. The parameters $\alpha$ and $\sigma$ characterize the velocity-weakening friction. The parameter $\sigma$ represents the difference between the maximum static friction 1 and the kinetic friction in the limit of $v=0$. The parameter $\alpha$ expresses the velocity-weakening rate of the kinetic friction, i.e., $d\phi(v)/dv=-2\alpha$ in the limit of $v=0$. When $v_i=dx_i/dt$ decreases to 0, the slip state changes to a stick state.  The stick state is assumed to be maintained until the force $k(x_{i+1}-2x_i+x_{i-1})+(v_0t-x_i)$ reaches the maximum static friction 1. It is assumed that the reverse motion does not occur in this model.~\cite{rf:4} We further assume that $v_0$ is infinitesimally small, and $v_0$ is set to be zero during the slip event. On the other hand, the pulling velocity $v_0$ is set to 1 in the stick state, which is equivalent to the measure of time with a unit of $1/v_0$. In our model, the starting time of the next slip event can be calculated from the maximum value of $k(x_{i+1}-2x_i+x_{i-1})+(v_0t-x_i)$ at the final time of each slip event. That is, the equation of motion Eq.~(1) is used only for slip processes.

We consider first the dynamics of a one-block system:
\begin{equation}
\frac{d^2x}{dt^2}=v_0t-x-\phi(v),
\end{equation}
where $v=dx/dt$. Figure 1(a) shows the time evolution of $x(t)$ for $\alpha=2$ and $\sigma=0.01$. The stick-slip motion repeats periodically. Slip size is defined as the slip distance $s=\Delta x=x(t_f)-x(t_i)$  between the initial time $t_i$ and the final time $t_f$ of each slip event. 
The slip size is 1.274 in the regular stick-slip process shown in Fig.~1(a).  
Figure 1(b) shows the relation of slip size as a function of $\alpha$ for $\sigma=0.01$. Slip size changes rapidly near $\alpha=1$. Figure 1(c) shows the relation of slip size as a function of $\alpha$ for $\sigma=0.0001$. The slip size is almost 0 for $\alpha<1$ and increases very rapidly near $\alpha=1$ like a phase transition. 

If $2\alpha v/(1-\sigma)$ is sufficiently small, Eq.~(3) with Eq.~(2) can be linearized as 
\begin{equation}
\frac{d^2x}{dt^2}=F-x-(1-\sigma)+2\alpha v,
\end{equation}
where $F=v_0t$. 
This is equivalent to the equation of motion of a linear amplified oscillation. To simplify the notation, $t-t_i$ is expressed as $t$ in the following part in this section. 
The initial condition is assumed to be $x(0)=0$ and $v(0)=0$. When the slip motion starts at $t=0$, the force $F-x(0)=F$ is equal to the maximum static force 1. The solution to Eq.~(4) for $\alpha<1$ is explicitly expressed as\cite{rf:21,rf:22}
\begin{equation}
x(t)=\sigma-\sigma e^{\alpha t}\left (\cos\sqrt{1-\alpha^2}t-\frac{\alpha}{\sqrt{1-\alpha^2}}\sin\sqrt{1-\alpha^2}t\right ),\;v(t)=\sigma e^{\alpha t}\frac{\sin\sqrt{1-\alpha^2}t}{\sqrt{1-\alpha^2}}.
\end{equation} 
At $t=\pi/\sqrt{1-\alpha^2}$, the velocity $v(t)$ becomes 0, and the slip motion stops. The slip size $s$ is evaluated at 
\begin{equation}
s=\sigma\left (1+e^{\alpha \pi/\sqrt{1-\alpha^2}}\right ).
\end{equation}
Slip size is proportional to $\sigma$ and diverges at $\alpha=1$. 
This relation is shown by the dashed line in Fig.~1(b) for $\sigma=0.01$. 
Good agreement is seen for a sufficiently small $\alpha$; however, the deviation becomes larger near $\alpha=1$. Slip size cannot be evaluated for $\alpha>1$ with the linear equation, because the nonlinearity is essential. This is related to the transition of the Carlson-Langer model near $\alpha=1$ suggested previously~\cite{rf:9,rf:21}.   
\begin{figure}[t]
\begin{center}
\includegraphics[height=4.cm]{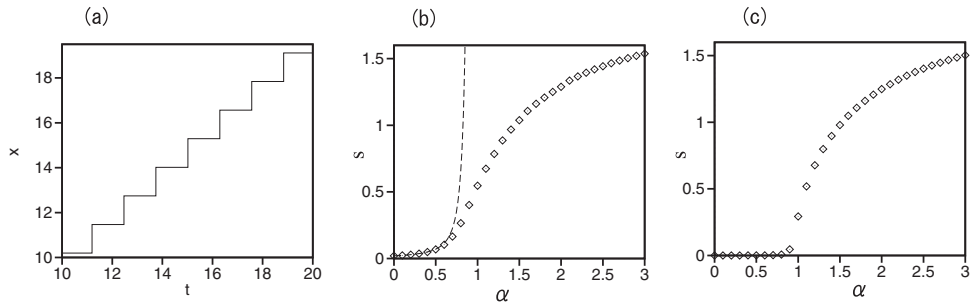}
\end{center}
\caption{Numerical results of the one-block system. (a) Time evolution of $x(t)$ at $\sigma=0.01$ and $\alpha=2$. (b) Slip size $s$ in a one-block model at $\sigma=0.01$. The dashed line is the curve obtained using Eq.~(6). (c) Slip size $s$ in a one-block model at $\sigma=0.0001$. }
\label{f1}
\end{figure}
\section{Intermittent Dynamics in Two-Block Systems}
Next, we study a two-block model:
\begin{eqnarray}
\frac{d^2x_1}{dt^2}&=&k(x_{2}-x_{1})+(v_0t-x_1)-\phi(v_1),\nonumber\\
\frac{d^2x_2}{dt^2}&=&k(x_{1}-x_{2})+(v_0t-x_2)-\phi(v_2).
\end{eqnarray}
The parameters $\sigma$ and $k$ are fixed to be $\sigma=0.01$ and $k=16$ in the following numerical simulations in Sects.~3-6.
Figure 2(a) shows the time evolution of the sum $S$ of the slip sizes $s_1$ and $s_2$ of the two blocks at $\alpha=2$. The initial condition is $x_1(0)=0.001,x_2(0)=0.002$, and $v_1(0)=v_2(0)=0$. For $t<42$, small slips of $S\sim 3.8\times 10^{-3}$ repeat several times ($3-20$ times), and then  a large slip of $S\sim 2.4$ occurs. 
Only one block moves in each small slip event. On the other hand, two blocks move together in the large slip event. This type of motion is called  a simultaneous slip in this paper.  A simultaneous slip occurs after several one-block slip processes, which we call an intermittent slip process.   
The sum $S$ in the large slip is roughly twice the slip size in the one-block system shown in Fig.~1(b). The intermittent slip process repeats in a chaotic manner.  Figure 2(b) shows the time evolution of the difference $x_2-x_1$ just after each slip event. For small slips, the difference $|x_2-x_1|$ decreases with time monotonically, and a large slip occurs when the difference becomes sufficiently small. If the difference $x_2-x_1$ is given just before the slip event, the difference $x_2^{\prime}-x_1^{\prime}$ just after the slip event can be numerically calculated using Eq.~(7). Figure 2(c) shows the relation of $x_2-x_1$ and $x_2^{\prime}-x_1^{\prime}$. For $x_2-x_1>0.0019$,  $x_2^{\prime}-x_1^{\prime}$ is approximated to be $x_2-x_1-0.0038$, and for $x_2-x_1<-0.0019$,  $x_2^{\prime}-x_1^{\prime}$ is approximated to be $x_2-x_1+0.0038$. 
The difference $|x_2-x_2|$ decreases by 0.0038 after the small slip process of the delayed block, and the delayed block approaches the other block. 
For a small slip process, an argument similar to Eqs.~(4)-(6) can be applied because only one block slips. The linear equation for the slipping block $i=1$ or 2 is given as
\[
\frac{d^2x_i}{dt^2}=F-(k+1)x_i-(1-\sigma)+2\alpha v_i,\]
 and  slip size can be evaluated as $s=\sigma/(k+1)\{1+\exp(\alpha \pi/\sqrt{1+k-\alpha^2})\}\sim 0.0039$. Note that the linear approximation is valid for the two-block system because $\alpha<k+1$ even for $\alpha>1$. 

When $|x_2-x_1|$ becomes smaller than 0.0019, the two blocks move together. If the two blocks move together and $x_1\sim x_2$ is satisfied, the coupling term $k(x_2-x_1)$ is nearly zero, and an essentially nonlinear slip process occurs for $\alpha>1$, as shown in Fig.~1(b) for the one-block system. 
The difference $x_2^{\prime}-x_1^{\prime}$ just after the simultaneous slip process depends strongly on the difference $x_2-x_1$ just before the slip event. The strong dependence is related to the complicated dynamics of the simultaneous slip motion.  
Figure 3(a) shows two time evolutions of $x_2(t)$ (dashed curves) and $x_1(t)$ (solid curves) for two initial conditions: $x_2(0)-x_1(0)=0.00142$  and 0.00169. Figure 3(b) shows the time evolution of the difference $x_2(t)-x_1(t)$ for the two initial values: $x_2(0)-x_1(0)=0.00142$ (solid curve) and 0.00169 (dashed curve). After the slip process, $x_2-x_1$ becomes negative for $x_2(0)-x_1(0)=0.00142$, that is, the position is reversed. The difference $\delta x=x_2-x_1$ exhibits an amplifying oscillation, because the linear approximation yields
\[\frac{d^2\delta x}{dt^2}=-(2k+1)\delta x-\frac{d\phi(v)}{dv}\frac{d\delta x}{dt},\]
where $d\phi(v)/dv<0$. 
\begin{figure}[t]
\begin{center}
\includegraphics[height=4.cm]{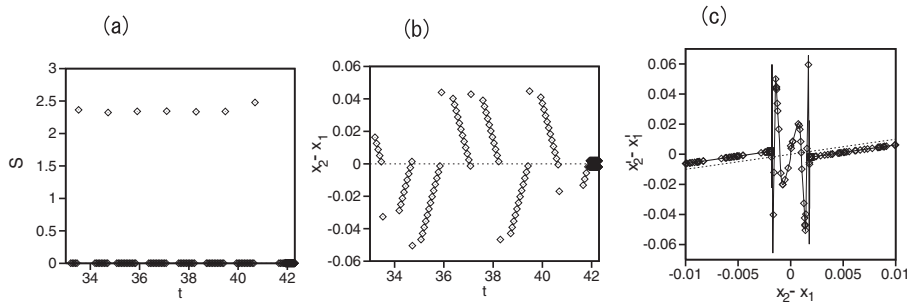}
\end{center}
\caption{Two-block system at $\sigma=0.01$, $k=16$, and $\alpha=2$. (a) Time evolution of the sum $S$ of the slip size of the two blocks. (b) Time evolution of the difference $x_2-x_1$. (c) Relation of $x_2-x_1$ vs $x_2^{\prime}-x_1^{\prime}$.}
\label{f2}
\end{figure}
\begin{figure}[t]
\begin{center}
\includegraphics[height=5.cm]{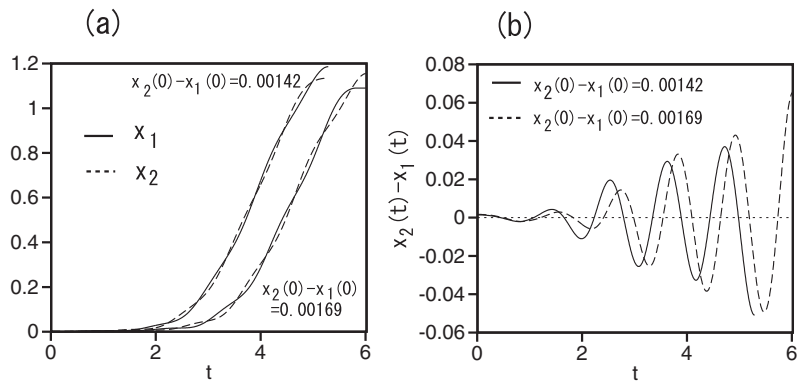}
\end{center}
\caption{Numerical results of the two-block system at $\sigma=0.01$, $k=16$, and $\alpha=2$. (a) Two time evolutions of $x_1(t)$ (solid curve) and $x_2(t)$ (dashed curve) for $x_2(0)-x_1(0)=0.00142$ and 0.00169 in a slip event. (b) Time evolutions of the difference $x_2(t)-x_1(t)$ for the two initial values $x_2(0)-x_1(0)=0.00142$ (solid curve) and 0.00169 (dashed curve). 
}
\label{f3}
\end{figure}
The one-dimensional map shown in Fig.~2(c) determines the dynamics of the stick-slip motion in the two-block system. 
The strong nonlinearity in the range $|x_2-x_1|<0.0019$ generates chaotic dynamics such as the logistic map.~\cite{rf:23} The maximum $|x_2^{\prime}-x_1^{\prime}|$ is 0.065, which is much larger than 0.00185. When a large $|x_2^{\prime}-x_1^{\prime}|$ is obtained by the one-dimensional mapping, small one-block slips continues many times. The sequence of small one-block slips corresponds to a laminar state in the intermittent time evolution. The duration of a laminar state after a large simultaneous slip is determined as $|x_2^{\prime}-x_1^{\prime}|/0.0038$ by the strong nonlinear mapping and the value of $|x_2-x_1|$ for the simultaneous slip.  

The intermittent time evolution repeats for a certain time; however, it is not an attractor. In Figs.~2(a) and 2(b), a stable two-period state appears for $t>42$. In the two-period state, a small slip repeats alternately for the two blocks, and the relation $x_2^{\prime}-x_1^{\prime}=-(x_2-x_1)=\pm 0.0019$ is satisfied.  That is, in a slip event, the delayed block slips while the other block is stuck, and the first block overtakes the second block. Then, in the next step, the second block slips while the first block is stuck. Small one-block slips repeat alternately for the first and second blocks. The two-period state is stable for a wide $\alpha$ range, and the chaotic state appears only transiently in most parameters. 

\begin{figure}[t]
\begin{center}
\includegraphics[height=4.cm]{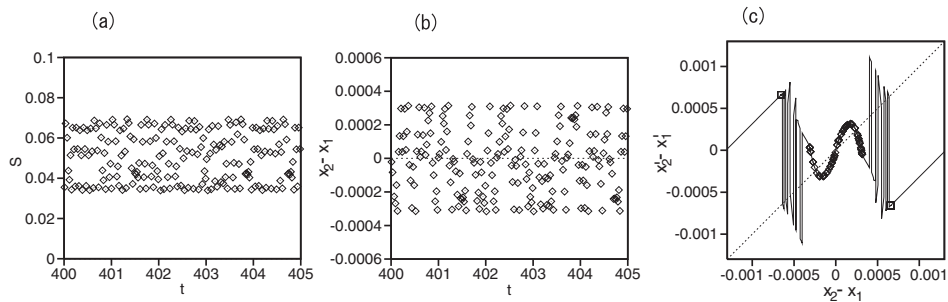}
\end{center}
\caption{Numerical results of the two-block system  at $\sigma=0.01$, $k=16$, and $\alpha=0.28$. (a) Time evolution of the sum $S$ of the slip size. (b) Time evolution of the difference $x_2-x_1$. (c) Relation of $x_2^{\prime}-x_1^{\prime}$ vs $x_2-x_1$.}
\label{f4}
\end{figure}
However, a chaotic state becomes stable for $0.25<\alpha<0.32$. 
Figure 4(a) shows the time evolution of the sum $S$ of the slip size for $\alpha=0.28$. Figure 4(b) shows the time evolution of the difference $x_2-x_1$. 
In this parameter range, the intermittent behavior is not observed. The two blocks move together or the simultaneous slips occur at each slip event.  
Figure 4(c) shows the relation of $x_2^{\prime}-x_1^{\prime}$ and $x_2-x_1$ at $\alpha=0.28$. 
The rhombi denote the result of the direct numerical simulation, and the curve is calculated using Eq.~(7). 
Chaotic dynamics appears in the central region where the two blocks slip together.  On the other hand, the two-period state is also stable. In the two-period state, the one-block slips occur alternately for the first and second blocks. In the two-period state, the relation $x_2^{\prime}-x_1^{\prime}=-(x_2-x_1)=\pm 0.00066$ is satisfied at $\alpha=0.28$. This relation is denoted by squares in Fig.~3(c). The two-block system is bistable at this $\alpha$. It consists of two states: an alternating-slip state and a simultaneous-slip state. Which state appears depends on the initial conditions.

Thus, we have found intermittent dynamics in two-block systems. That is, small slips make the difference $|x_1-x_2|$ decrease, a large simultaneous slip occurs when the difference is sufficiently small, and the difference is amplified by a large simultaneous slip.  Chaotic behaviors were already observed in asymmetric two-block systems previously.~\cite{rf:24,rf:25} However, the intermittent behavior of a large slip after the repetition of several small slips for $\alpha>1$ was not so emphasized in previous papers.    

\section{Intermittent Dynamics in One-Dimensional Systems}
In this section, we investigate intermittent dynamics in one-dimensional systems expressed by Eq.~(1). 
The total block number is denoted as $N$, and the periodic boundary conditions are imposed in the numerical simulation. 
Spatio-temporal chaos is observed for most initial conditions; however, there is a stable traveling wave solution even for a large $N$. 
The regular traveling wave solution corresponds to the two-period state in  two-block systems, in that only one block slips in each slip event, and the neighboring block slips in the next slip event. We first show a regular traveling wave for $N=200$. (The traveling wave state was also obtained in a smaller system such as $N=10$.) The parameters $k$ and $\sigma$ are fixed to be $k=16$ and $\sigma=0.01$. 
Figure 5 shows such a traveling wave state at $\alpha=1.5$. Only one block slips at each slip event, and the slip event propagates in the right direction. The slip size is $1.007\times 10^{-3}$, which is close to the approximate value of the one-block slip: $s=\sigma/(2k+1)\{1+\exp(\alpha \pi/\sqrt{1+2k-\alpha^2})\}\sim 1.03\times 10^{-3}$. The traveling wave state in Fig.~5 was obtained using a special method by gradually increasing $\alpha$ from $\alpha=0$ for $N=200$.  This is because such a traveling wave state is hardly obtained under most initial conditions in the case of a large $N$ at a fixed large $\alpha$. (An intermittent chaotic state is obtained under most initial conditions.) The traveling wave solution is difficult to obtain, but it is an attractor in this dynamical system. 

\begin{figure}[t]
\begin{center}
\includegraphics[height=3.5cm]{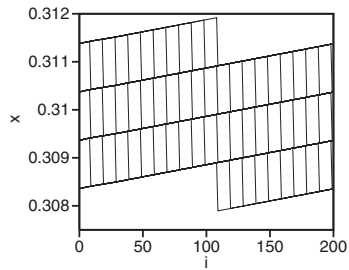}
\end{center}
\caption{Traveling wave state in the one-dimensional system at $\sigma=0.01,k=16, \alpha=1.5, k=16$, and $N=200$.}
\label{f5}
\end{figure}
\begin{figure}[t]
\begin{center}
\includegraphics[height=4.cm]{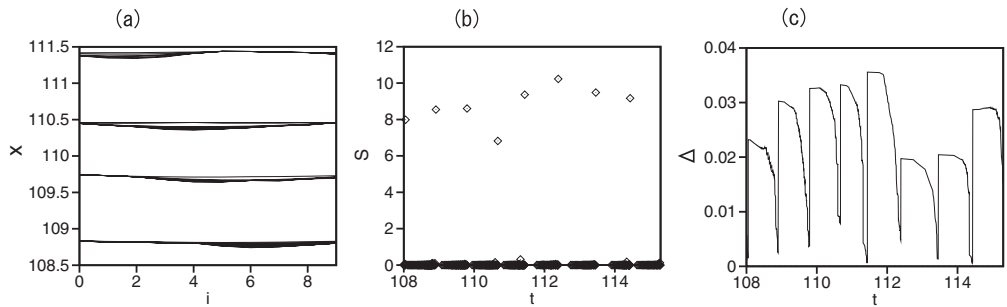}
\end{center}
\caption{Numerical results of the one-dimensional system at $\sigma=0.01,k=16,\alpha=1.5$, and $N=10$. (a) Time evolution of the profile of $x_i(t)$. (b) Time evolution of $S(t)$ in the chaotic state for $108<t<115.3$. (c) Time evolution of $\Delta$ for $108<t<115.3$.}
\label{f6}
\end{figure}
\begin{figure}[tbp]
\begin{center}
\includegraphics[height=4.cm]{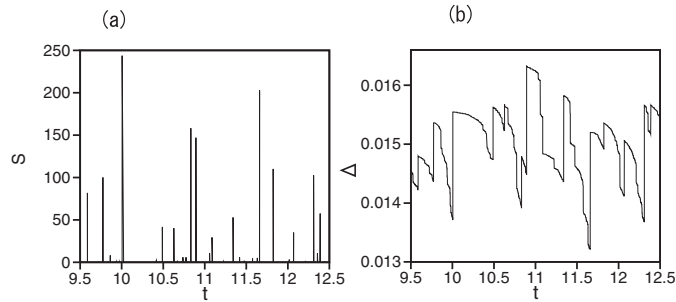}
\end{center}
\caption{Numerical results of the one-dimensional system at  $\sigma=0.01,k=16,\alpha=1.5$, and $N=500$. (a) Time evolution of $S(t)$ for $9.5<t<12.5$. (b) Time evolution of $\Delta $ for $9.5<t<12.5$.}
\label{f7}
\end{figure}
The intermittent dynamics is easily observed even in a system of small $N$ for a large $\alpha$. 
Figure 6(a) shows the time evolution of the profile of $x_i(t)$ in the chaotic state at $\alpha=1.5$ and $N=10$. Total slip size is defined as
\begin{equation}
S(t)=\sum_{i=1}^{N}s_i,
\end{equation}
where $s_i=\Delta x_i=x_i(t_f)-x_i(t_i)$ is the distance of the slip for the $i$th block at each slip event between $t_i$ and $t_f$. 
Figure 6(b) shows the time evolution of the sum $S(t)$ at $\alpha=1.5$ and $N=10$. There is a dent structure in Fig.~6(a). The delayed blocks are located near the bottom of the dent. 
Small slips of one or a few blocks occur many times for the delayed blocks and the dent profile of $x_i$ gradually becomes flat. The smallest slip is $1.007\times 10^{-3}$ for a single-block slip. 
After a sequence of small slip events, a large slip, i.e., a simultaneous slip, occurs once. 
The spatial irregularity of the profile of $x_i$ can be measured by the quantity $\Delta$, which is defined by
\begin{equation}
\Delta=(1/N)\sum_{i=1}^N|x_{i+1}-x_i|.
\end{equation}  
Figure 6(c) shows the time evolution of $\Delta$. The irregularity $\Delta$ decreases with time during the stage of many small slips, and $\Delta$ jumps to a large value when a large slip occurs. 
That is, the repetition of small slips sweeps out the irregularity (dent structure) of $x_i$ and prepares the occurrence of a large slip. If the profile of $x_i$  is completely flat, it is natural that all the blocks slip simultaneously. In our ten-block system, a simultaneous slip occurs when the dent structure of $x_i$ becomes very shallow, as shown in Fig.~6(a).  
The duration of large slips is relatively long, and therefore, the difference $x_i-x_j$ is amplified during a large slip, and a deep dent structure is created again. On the rough profile of $x_i$, only small slips occur.  Thus, the intermittent slip process repeats many times. It is a mechanism of the spatio-temporal chaos in our block-spring system in the case of $\alpha>1$. There is only one dent in the system of $N=10$. However, a complex structure including many dents appears for a large $N$. 
Figure 7 shows the numerical results in a larger system of $N=500$ at $\alpha=1.5$ and $k=16$. Figure 7(a) shows the time evolution of $S(t)$. Large slips occur intermittently. Many small slips occur between large-size slips but are hardly visible in this plot. 
Figure 7(b) shows the time evolution of $\Delta$. $\Delta$ jumps upward when a large slip of $S>30$ occurs, and $\Delta$ decreases during the repetition of small slips. This confirms the conjecture that the spatial irregularity decreases with small slips, a large slip occurs on a relatively smooth profile of $x_i$, and this large slip increases the spatial irregularity. 

Thus, we have interpreted the complex intermittent dynamics of the one-dimensional Carlson-Langer model on the basis of the intermittent dynamics of two-block systems and the quantity $\Delta$.    
 
\section{Magnitude Distribution in  One- and Two-Dimensional Systems}
In this section, we explain the magnitude distribution and show the numerical results in the one- and two-dimensional systems as an introduction to the next section.  
In this paper, magnitude is defined as 
\begin{equation}
M=\ln S,
\end{equation}
where $S$ is the sum of the slip sizes at all blocks. 
$M$ corresponds to the moment magnitude in seismology. We perform a long-time numerical simulation of a one-dimensional system and construct the distribution of $M$ from the long-time evolution of $S$. The parameters $k$ and $\sigma$ are fixed to be $k=16$ and $\sigma=0.01$. 
Figure 8(a) shows the magnitude distribution $P(M)$ at $\alpha=2$ and $N=2000$. A hump structure is observed near $M\sim 4$ at $\alpha=2$, which was the same structure observed in the original studies.~\cite{rf:4,rf:5} The hump structure for a large $\alpha$ is directly related to the magnitude of a large slip, which is close to the number of slipped blocks times the slip size for a one-block system expressed by Eq.~(6). A large slip occurs intermittently after a sequence of small slips even in this large system. This corresponds to the intermittent behavior found in the initial stage for the two-block system and a system of $N=10$ for a large $\alpha$ satisfying $\alpha>1$.  Such a large-scale slip hardly occurs at $\alpha=0.25<1$, because the slip size is on the order of $S\sim N \sigma(1+e^{\alpha\pi/\sqrt{1-\alpha^2}})$ even if $N$ blocks slip simultaneously for $\alpha<<1$. As a result, the magnitude distribution decays rapidly near $M=1$. 

An exponential behavior is observed at an intermediate $\alpha$, i.e., $\alpha\sim 1$. The exponential distribution is expressed as
\begin{equation}
P(M)\sim e^{-bM}.
\end{equation}
The exponent $b$ is evaluated to be 0.45 at $\alpha=1$ in this one-dimensional model. It is a fitting between $M=-6$ and $M=3$, and the error is evaluated to be $\pm 0.02$. 

The probability distribution of $S$ is expressed as 
\begin{equation}
P(S)\sim S^{-b-1}
\end{equation}
from the relation $M=\ln S$. That is, the slip-size distribution obeys a power law. 
The Gutenberg-Richter law is usually expressed as the cumulative distribution $N(S_0)=\int_{S_0}^{\infty}P(S)dS\sim S_0^{-b}$.  $b$ is evaluated to be $2/3$ using the data of all the earthquakes in the world.~\cite{rf:2} 

\begin{figure}
\begin{center}
\includegraphics[height=4.cm]{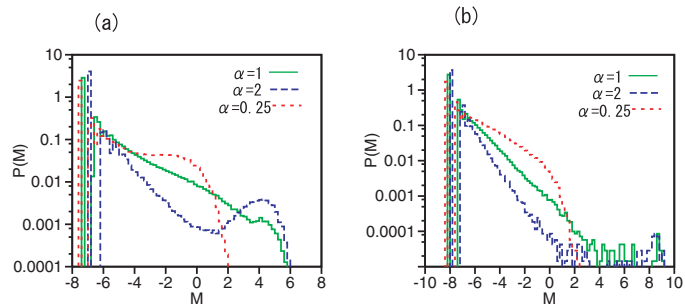}
\end{center}
\caption{(Color online) (a) Magnitude distributions $P(M)$ at $\alpha=2,1$, and 0.25 in a one-dimensional system of $N=2000$ at  $\sigma=0.01$ and $k=16$. (b) Magnitude distributions for a two-dimensional model on a square lattice of $150\times 150$ at $\alpha=2,1$, and 0.25 for $\sigma=0.01$ and $k=16$. }
\label{f8}
\end{figure}
The two-dimensional Carlson-Langer model on a square lattice is expressed as 
\begin{equation}
\frac{d^2x_{i,j}}{dt^2}=k(x_{i+1,j}+x_{i-1,j}+x_{i,j+1}+x_{i,j-1}-4x_{i,j})+(v_0t-x_{i,j})-\phi(v_{i,j}),
\end{equation}
where $k$ is the coupling constant with the neighboring block. 
We have performed numerical simulation of the two-dimensional model on a square lattice of $150\times 150$.  Figure 8(b) shows the magnitude distribution in the Carlson-Langer model on the square lattice at $\alpha=2,1$, and 0.25. The parameters $k$ and $\sigma$ are fixed to be $k=16$ and $\sigma=0.01$. 
There is a small peak structure near $M=8.5$ for $\alpha=2$. 
This implies that very large slip events occur once in a while. The magnitude distribution also decays rapidly near $M=1$ at $\alpha=0.25$ in this two-dimensional system.  
A power-law behavior for the slip size is observed at $\alpha=1$. The exponent $b$ is evaluated as 0.85 at $\alpha=1$. It is a fitting between $M=-7$ and $M=4$, and the error is evaluated as $\pm 0.02$. 

\begin{figure}[t]
\begin{center}
\includegraphics[height=4.cm]{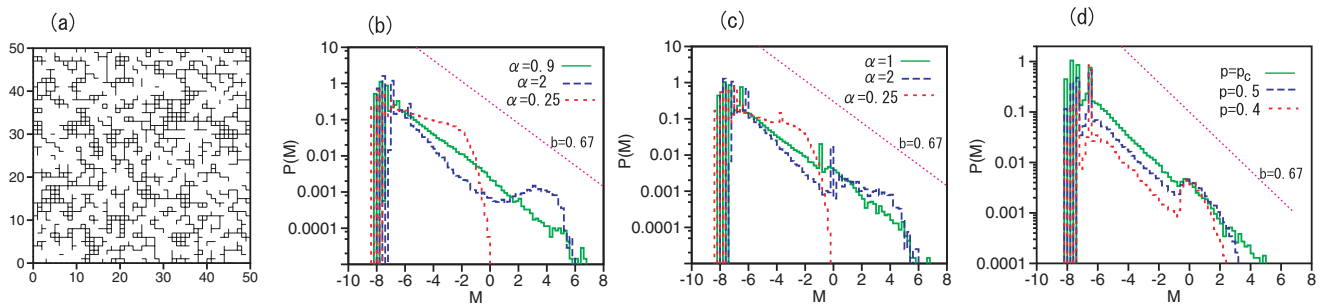}
\end{center}
\caption{(Color online) (a) Percolation clusters at $p=0.56$ in a system of $50\times 50$. (b) Magnitude distributions in a two-dimensional system of $150\times 150$ on the critical percolation cluster at $\alpha=2,0.9$, and 0.25. The parameters $\sigma$ and $k$ are set to be $\sigma=0.01$ and $k=16$. The straight dashed line denotes the line of $b=0.67$. (c) Magnitude distribution in a two-dimensional model on the whole clusters of the critical site percolation at $\alpha=2,1$, and 0.25, and $\sigma=0.01$ and $k=16$. The straight dashed line denotes the line of $b=0.67$. (d) Magnitude distribution for the percolation clusters for $p=p_c,0.5$, and 0.4 at $\alpha=1$, $\sigma=0.01$, and $k=16$. The straight dashed line denotes the line of $b=0.67$.}
\label{f9}
\end{figure}
\begin{figure}[t]
\begin{center}
\includegraphics[height=4.cm]{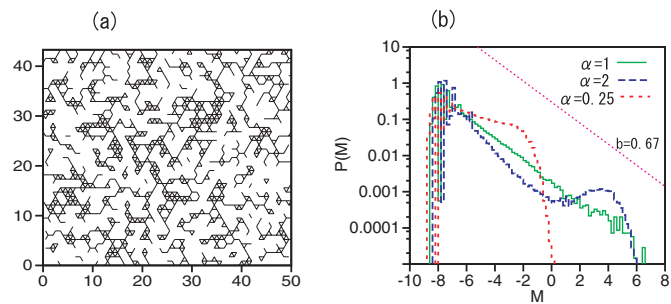}
\end{center}
\caption{(Color online) (a) Percolation clusters at $p=0.5$ on a triangular lattice. (b) Magnitude distribution in a two-dimensional system of $150\times 150$ on the critical percolation cluster of the triangular lattice at $\alpha=2,1$, and 0.25. The parameters $\sigma$ and $k$ are  $\sigma=0.01$ and $k=16$. The straight dashed line denotes the line of $b=0.67$.}
\label{f10}
\end{figure}
\section{Carlson-Langer Model on Percolation Clusters}
Various sizes of asperities are considered to exist at the boundary of two plates of plate tectonics in the theory of seismology, and asperities play an important role in determining the magnitude of earthquakes. In this study, we use percolation clusters to study the effect of such heterogeneous contact of two plates.  In the site percolation model, a block site is selected with probability $p$ on the square lattice. If there is a block in the neighboring site of the selected block site, the two blocks are coupled with a spring of spring constant $k$. 
That is, we consider that the two plates are in contact with each other via the selected block site. A critical percolation is obtained at $p=p_c=0.5927$ on a square lattice and the critical percolation cluster is fractal. Figure 9(a) shows percolation clusters at $p=0.56<p_c$ in a system of $50\times 50$. In Fig.~9(a), the bonds coupled with springs between the selected blocks are plotted. 

We perform a numerical simulation of the Carlson-Langer model on percolation clusters to investigate the geometrical and inhomogeneity effects on slip dynamics. We perform three kinds of numerical simulation, that is,  on the largest percolation cluster, on the whole clusters at the critical value of $p=p_c$,  and on the whole clusters at several subcritical values satisfying $p<p_c$. The system size is $150\times 150$. 
The parameters $k$ and $\sigma$ are fixed to be $k=16$ and $\sigma=0.01$. 
Figure 9(b) shows the magnitude distribution for the single largest cluster. The magnitude distributions at $\alpha=2,0.9,$ and 0.25 are shown in Fig.~9(b). A hump structure is observed near $M=3$ for $\alpha=2$. The magnitude distribution decays rapidly near $M=-1$ at $\alpha=0.25$  in this critical percolation cluster. A power law behavior for the slip size is observed at $\alpha=0.9$. The exponent $b$ is nearly 0.67, which is close to the exponent of the Gutenberg-Richter law of earthquakes. The transition seems to be continuous. That is, the hump structure appears gradually for $\alpha>1.1$ and  the rapid decay at a large $M$ becomes clear for $\alpha<0.7$.  These behaviors are similar to those observed in the one-dimensional system, except that $b$ value is different.  

Figure 9(c) shows the magnitude distributions at $\alpha=2,1$, and 0.25 for the whole clusters at $p=p_c$. There are many clusters in this system and  the cluster size obeys a power law at $p=p_c$. The magnitude distribution has a singular peak, which corresponds to slip events for independent blocks. A single block of cluster size 1 exhibits a periodic stick-slip motion, and slip size is uniquely determined as shown in Fig.~1(b), which contributes to the singular peak in Fig.~9(c).  A shoulder structure is observed in the magnitude distribution at $\alpha=2$ near $M\sim 2$, probably because the hump structure smears out owing to the size distribution of clusters. A power law behavior with the exponent $b=0.67$ is observed at $\alpha=1$ even in this many-cluster system. 
The magnitude distribution decays rapidly near $M=-1$ at $\alpha=0.25$. 

Figure 9(d) shows the magnitude distribution for $\alpha=1$ at $p=p_c,0.5$, and 0.4. In this numerical simulation, we have removed independent blocks. As a result, the singular peak in the magnitude distribution disappears.  
A power-law behavior with the exponent $b\sim 0.67$ is observed at $p=p_c$. 
A small peak appears near $M\sim 0$ for $p=0.5$ and 0.4. 
The magnitude distribution decays rapidly at $p=0.4$ for $M>1$, probably because there are no sufficiently large clusters. In other words, the power law behavior becomes unclear as $p_c-p$ becomes larger. That is, the power law distribution of clusters is an important factor for the power law of slip size.  

We further study the Carlson-Langer model in a critical percolation cluster on a triangular lattice. The critical percolation appears at $p=p_c=0.5$ on the triangular lattice for the site percolation. Figure 10(a) shows percolation clusters on the triangular lattice at $p=0.5$. We perform a  numerical simulation on a critical percolation cluster at $p=0.5$. The parameters $k$ and $\sigma$ are fixed to be $k=16$ and $\sigma=0.01$, and the system size is $150\times 150$. Figure 10(b) shows the magnitude distribution at $\alpha=2,1$, and 0.25. 
A power law behavior is observed at $\alpha=1$, and the exponent $b$ is evaluated as $b=0.67$. These results suggest that  $b$ does not depend on the substrate lattice structure. 

\section{Summary and Discussion}
The Carlson-Langer model of earthquakes has been studied by many authors. However, its complicated intermittent dynamics remains to be completely understood, and the effect of the asperity was not investigated in the model. We have numerically investigated the mechanism of the spatio-temporal intermittency for $\alpha>1$ and the effect of asperity on the magnitude distribution using percolation clusters. 

We have shown that the intermittent behavior appears in a two-block system for $\alpha>1$, and explained the behavior using a one-dimensional map. We have proposed a simplified viewpoint for the intermittent spatio-temporal chaos in a larger one-dimensional system at $\alpha>1$, using the result of the two-block system and the quantity $\Delta$ expressing spatial irregularity. That is, a sequence of small slips makes the profile of $x_i$ uniform,  and then a large slip occurs on a nearly flat profile of $x_i$. The duration of the large slip is long and the spatial irregularity of $x_i$ is amplified during a long simultaneous slip. After the large slip, small slips occur again in the rough profile of $x_i$.
 It might be a new viewpoint that the small slips actively prepare the condition s for a large slip, because slips of all sizes are usually considered to reduce the stress, decreasing the probability of large earthquakes.  

We have checked that a power-law behavior in the slip size distribution appears at the intermediate $\alpha\sim 1$ in the one- and two-dimensional systems. Then, we have performed a numerical simulation on percolation clusters on the square lattice, taking the concept of asperities of various sizes into consideration. The magnitude distribution of the total slip-size on percolation clusters obeys a power law of the exponent $b\sim 2/3$ near $\alpha=1$ at the critical point of $p=p_c$.  The range of the power law is reduced as $p$ deviates from $p_c$. This might be related to the fractal asperity hypothesis. We have checked that a similar power law is obtained for a percolation cluster on a triangular lattice.

We have obtained a power law of $b\sim 2/3$ in the Carlson-Langer model on percolation clusters; however, we are not sure whether the power law is directly related to the  Gutenberg-Richter law, because the power law of $b\sim 2/3$ appears only under special conditions: $\alpha\sim 1$ and $p\sim p_c$, and there is no reason for these special values in nature. 
 
There are controversies regarding the self-organized criticality theory vs the fractal asperity theory for explaining the origin of the Gutenberg-Richter law. 
Our numerical results of the Carlson-Langer model do not directly support either theory.  That is, a special parameter $\alpha$ needs to be chosen for the power laws, which does not support the theory of self-organized criticality. The fractal distribution at $p=p_c$ does not always lead to a power law, which does not support the fractal asperity theory.  
Our model might be in an intermediate of the two theories, in that spatial irregularity is self-organized by the intermittent dynamics of the deterministic model, and that the fractal asperity can be included in the model.  

\end{document}